\documentclass{aastex}
\usepackage{emulateapj5}
\usepackage{apjfonts}
\usepackage{epsf}

\shorttitle{Fermi acceleration at relativistic shocks}
\shortauthors{Lemoine \& Pelletier}

\begin{document}


\title{Particle Transport in Tangled Magnetic Fields and Fermi Acceleration at
  Relativistic Shocks}


\author{Martin Lemoine\altaffilmark{1}}
\affil{GReCO / Institut d'Astrophysique de Paris, CNRS,\\
98 bis boulevard Arago, F-75014 Paris, France}
\altaffiltext{1}{email: lemoine@iap.fr}
%\and
\author{Guy Pelletier\altaffilmark{2}}
\affil{Laboratoire d'Astrophysique de Grenoble LAOG,\\
CNRS, Universit\'e Joseph Fourier,\\
BP~53, F-38041 Grenoble, France,\\
Institut Universitaire de France}
\altaffiltext{2}{email: guy.pelletier@obs.ujf-grenoble.fr}


%\altaffiltext{1}{Visiting Astronomer, Cerro Tololo Inter-American Observatory.
%CTIO is operated by AURA, Inc.\ under contract to the National Science
%\altaffiltext{2}{Society of Fellows, Harvard University.}
%\altaffiltext{3}{present address: Center for Astrophysics,
%    60 Garden Street, Cambridge, MA 02138}
%\altaffiltext{4}{Visiting Programmer, Space Telescope Science Institute}
%\altaffiltext{5}{Patron, Alonso's Bar and Grill}


\begin{abstract}
This paper presents a new method of Monte-Carlo simulations of test
particle Fermi acceleration at relativistic shocks. The particle
trajectories in tangled magnetic fields are integrated out exactly
from entry to exit through the shock, and the conditional probability
of return as a function of ingress and egress pitch angles is
constructed by Monte-Carlo iteration. These upstream and downstream
probability laws are then used in conjunction with the energy gain
formula at shock crossing to reproduce Fermi acceleration.  For pure
Kolmogorov magnetic turbulence upstream and downstream, the spectral
index is found to evolve smoothly from $s=2.09\pm0.02$ for mildly
relativistic shocks with Lorentz factor $\Gamma_{\rm s}=2$ to $s\simeq
2.26\pm0.04$ in the ultra-relativistic limit $\Gamma_{\rm s}\gg1$.
The energy gain is $\sim \Gamma_{\rm s}^2$ at first shock crossing,
and $\sim 2$ in all subsequent cycles as anticipated by Gallant \&
Achterberg (1999). The acceleration timescale is found to be as short
as a fraction of Larmor time when $\Gamma_{\rm s}\gg1$.

\end{abstract}




\section{Introduction}

Fermi acceleration at relativistic shocks is an important topic for
understanding the formation of spectra of ultrarelativistic particles
and radiation in relativistic flows such as those observed in active
nuclei, microquasars, $\gamma-$ray bursts and pulsar wind nebulae (see
Kirk \& Duffy 2001 and references therein).  Of particular interest is
the acceleration timescale that can be as short as a Larmor time for
relativistic Fermi acceleration; the smaller return probability to the
shock for downstream particles, as compared to the non-relativistic
regime, is compensated by the much larger energy gain at each
cycle. However the study of Fermi acceleration in the relativistic
limit is more involved than in the non-relativistic regime due to the
increased importance of the anisotropy of the distribution function
(Gallant \& Achterberg 1999).

Various methods have been used to study the relativistic regime of
Fermi acceleration (see Kirk \& Duffy 2001 and references therein),
starting with analytical estimates by Peacock (1981), followed by
semi-analytical methods (Kirk \& Schneider 1987; Gallant \& Achterberg
1999; Kirk et al. 2000; Achterberg et al. 2001) and numerical
Monte-Carlo techniques (Ballard \& Heavens 1992; Ostrowski 1993;
Bednarz \& Ostrowski 1998), which in spite of their differences
converge to an asymptotic spectral index $s\approx 2.2-2.3$ in the
ultra-relativistic limit.

In the present {\it Letter}, we propose a new numerical Monte-Carlo
method of simulation of the acceleration of test particles at
relativistic shocks. The trajectories of particles in the upstream and
downstream inhomogeneous magnetic fields are integrated out exactly
from the entry of each particle through the shock until its first
return to the shock. The law of probability of return to the shock as
a function of ingress and egress pitch angles is then constructed by
Monte-Carlo iteration. Finally we combine these probability laws, one
defined for upstream and the other for downstream, with the energy
gain formula at shock crossing to simulate the acceleration
process. This latter use of the angular probability laws is similar to
the method of Gallant et al. (2000) with some differences to be
discussed below.

   The present method appears efficient and potentially more powerful
when compared to direct Monte-Carlo simulations which follow each
particle through its repeated shock crossings (e.g. Ballard \& Heavens
1992; Ostrowski 1993). It allows one to simulate relativistic Fermi
acceleration in any magnetic configuration, albeit for test particles
only. We describe the method and numerical simulations in Section~2
and then present the results for a planar shock with fully turbulent
magnetic field both upstream and downstream in Section~3.

\section{Numerical simulations}

The hydrodynamic jump conditions at an adiabatic strong shock,
neglecting magnetic fields, are given in Blandford \& Mc~Kee (1977),
Kirk \& Duffy (2001) and Gallant (2002). The shock Lorentz factor is
$\Gamma_{\rm s}$ (upstream or lab frame\footnote{Unless otherwise
noted, all quantities are calculated in the upstream (lab) frame;
wherever needed, quantities relative to a given frame but calculated
in an other will be marked with the subscript $\vert$, e.g.,
$\beta_{\rm s\vert d}$ refers to the shock velocity measured in the
downstream rest frame.}), and the relative Lorentz factor $\Gamma_{\rm
r}$ between upstream and downstream $\Gamma_{\rm r}\equiv \Gamma_{\rm
s} \Gamma_{\rm s\vert d}(1-\beta_{\rm s}\beta_{\rm s\vert d})$. The
downstream Lorentz factor $\Gamma_{\rm s\vert d}$ (and $\Gamma_{\rm r}$) 
can be obtained as a
function of $\Gamma_{\rm s}$ from the relations derived from the shock
jump conditions given in Gallant (2002).
%\begin{equation}
%  \Gamma_{\rm s\vert d}^2 \,  =  \, \frac{F(\xi)}{F(\xi)-1}\, , \quad
%  \Gamma_{\rm s}^2 \,  =  \, G(\xi)^2 \frac{F(\xi)}{F(\xi)-1}\
%     \label{eq:BDE}
%\end{equation}
%where $\xi \equiv mc^{2}/T_{\rm d\vert d}$, $T_{\rm d\vert d}$ being
%the downstream temperature and $m$ the particle mass, $G(\xi) \equiv
%K_{3}(\xi)/K_{2}(\xi)$, with $K_2$, $K_3$ modified Bessel functions,
%and $F(\xi)\equiv [\xi G(\xi) - 1]^2 - \xi^2$. These relations hold
%for a gas composed of possibly different particles species but with
%same $\xi$ (Gallant 2002). Equations~(\ref{eq:BDE}) can be inverted
%numerically to obtain $\Gamma_{\rm s \vert d}$ as a function of
%$\Gamma_{\rm s}$.  
In particular, in the ultra-relativistic limit $\Gamma_{\rm
s}\rightarrow +\infty$, one finds the well-known results
$\beta_{\rm s\vert d}\rightarrow 1/3$ and $\Gamma_{\rm r}\rightarrow
\Gamma_{\rm s}/\sqrt{2}$.

We conduct our simulations in two steps. We first perform Monte-Carlo
simulations of particle propagation in a given magnetic field
structure following Casse, Lemoine \& Pelletier (2001) (wherein one
may find more details on the numerical procedure). These simulations
are carried out separately in the downstream and in the upstream rest
frames. It is possible to set up any magnetic field structure
including regular and tangled components, but in the following, for
the sake of simplicity, we restrict ourselves to the case of pure
Kolmogorov turbulence in both downstream and upstream rest frames. The
equations of motion of each particle are integrated out exactly, the
magnetic field being calculated at each point of the trajectory as a
sum of plane waves, using 200 modes spaced logarithmically on three
decades of wavelength and whose wavevector directions are drawn at
random\footnote{Using a higher number of modes does not modify the
results shown here as we have checked.}. Ostrowski (1993) used a
similar technique to construct the magnetic field albeit with 3 modes
only, while Ballard \& Heavens (1992) used three-dimensional FFT
methods (see Casse, Lemoine \& Pelletier 2001 for a comparison of
these methods).

   The trajectories are integrated over 100 scattering times upstream
and 1000 scattering times downstream in order not to miss possible
late returns to the shocks.  The laws of return probability as a
function of ingress and egress pitch angles are then constructed in
the following way. We draw at random a point along a simulated
trajectory which defines the point of entry through the shock. The
ingress pitch angle cosine to shock normal $\mu^{\rm i}$ is recorded,
the trajectory scanned to find the point of exit through the shock,
and the egress pitch angle cosine $\mu^{\rm e}$ is then
recorded. Iteration of the above then yields the desired law of
conditional return probability ${\cal P}(\mu^{\rm i};\mu^{\rm e})$
which gives the probability for a particle entering with a pitch angle
cosine $\mu^{\rm i}$ to return to the shock with a pitch angle cosine
$\mu^{\rm e}$. The simulations also give a direct measurement of the
return time to the shock as a function of pitch angles.

Once the upstream and downstream laws of return probability,
respectively ${\cal P}_{\rm u}\left(\mu_{\rm u}^{\rm i}; \mu_{\rm
u}^{\rm e}\right)$ and ${\cal P}_{\rm d}\left(\mu_{\rm d}^{\rm i};
\mu_{\rm d}^{\rm e}\right)$ are known, the simulation of the
acceleration process itself can be performed as follows. We denote by
$f_{\rm d}^{2n+1}(\mu_{\rm d},\epsilon_{\rm d})$ the distribution
function of particles that enter the shock to downstream and that have
experienced $2n+1$ shock crossings, and $f_{\rm u}^{2n}(\mu_{\rm
u},\epsilon_{\rm u})$ similarly upstream: particles are upstream for
an even number of shock crossings and $f_{\rm u}^0$ represents the
injection population.  These distribution functions are normalized to
the total number of particles $N$ injected such that, in the absence
of escape from the acceleration site, after $2n$ shock crossings
$N=\int{\rm d}\mu_{\rm u}{\rm d}\epsilon_{\rm u}f^{2n}_{\rm
u}(\mu_{\rm u},\epsilon_{\rm u})$, and after $2n+1$ shock crossings
$N=\int{\rm d}\mu_{\rm d}{\rm d}\epsilon_{\rm d}f^{2n+1}_{\rm
d}(\mu_{\rm d},\epsilon_{\rm d})$. The conservation of particle number
at shock crossing $u\rightarrow d$ and Lorentz transforms of pitch
angles and energies imply:

\begin{equation}
f_{\rm d}^{2n+1}(\epsilon_{\rm d}, \mu_{\rm d}^{\rm i}) {\rm d}\mu_{\rm
d}^{\rm i}{\rm d}\epsilon_{\rm d} \, = \, \left[ \int_{\beta_{\rm
s}}^{1}{\rm d}\mu_{\rm u}^{\rm i}\,{\cal P}_{\rm u}(\mu_{\rm u}^{\rm
i}; \mu_{\rm u}^{\rm e}) f_{\rm u}^{2n}(\epsilon_{\rm u};\mu_{\rm
u}^{\rm i})\right] {\rm d}\mu_{\rm u}^{\rm e}{\rm d}\epsilon_{\rm u}
\label{eq:mapud_f}
\end{equation}
\begin{equation}
\mu_{\rm d}^{\rm i} \, = \, \frac{\mu_{\rm u}^{\rm e} - \beta_{\rm
r}}{1-\beta_{\rm r}\mu_{\rm u}^{\rm e}} ,\quad \epsilon_{\rm d} \, =
\, \Gamma_{\rm r}(1-\beta_{\rm r}\mu^{\rm e}_{\rm u})\epsilon_{\rm u}
,
\label{eq:mapud_L}
\end{equation}
and one obtains a similar system for shock crossing $d\rightarrow u$:
\begin{equation}
f_{\rm u}^{2n}(\epsilon_{\rm u}, \mu_{\rm u}^{i}) {\rm d}\mu_{\rm
u}^{\rm i}{\rm d}\epsilon_{\rm u} \, = \, \left[ \int_{-1}^{\beta_{\rm
s\vert d}}{\rm d}\tilde\mu_{\rm d}^{\rm i}\,{\cal P}_{\rm d}(\tilde\mu_{\rm
d}^{\rm i}; \mu_{\rm d}^{\rm e}) f_{\rm d}^{2n-1}(\tilde\epsilon_{\rm
d};\tilde\mu_{\rm d}^{\rm i})\right] {\rm d}\mu_{\rm d}^{\rm e}{\rm
d}\tilde\epsilon_{\rm d}
\label{eq:mapdu_f}
\end{equation}
\begin{equation}
\mu_{\rm u}^{\rm i} \, = \, \frac{\mu_{\rm d}^{\rm e} + \beta_{\rm
r}}{1+\beta_{\rm r}\mu_{\rm d}^{\rm e}}, \quad \epsilon_{\rm u} \, =
\, \Gamma_{\rm r}(1+\beta_{\rm r}\mu^{\rm e}_{\rm d})\tilde\epsilon_{\rm d}
,
\label{eq:mapdu_L}
\end{equation}

The terms within brackets in Eqs.~(\ref{eq:mapud_f}) and
(\ref{eq:mapdu_f}) correspond to the distribution function upon exit
from upstream and downstream respectively. The return probability to
the shock $P_{\rm ret}(\mu^{\rm i}_{\rm d})$ as a function of ingress
pitch angle can be obtained as: $P_{\rm ret}(\mu^{\rm i}_{\rm
d})\equiv \int {\rm d}\mu^{\rm e}_{\rm d}\, {\cal P}_{\rm d}(\mu^{\rm
i}_{\rm d};\mu^{\rm e}_{\rm d})$. The corresponding return probability
for upstream is obviously unity. After each cycle $u\rightarrow
d\rightarrow u$, a fraction $f_{\rm out}^{2n+1}(\epsilon)=\int {\rm
d}\mu_{\rm d}^{\rm i} [1- P_{\rm ret}(\mu_{\rm d}^{\rm i})]f_{\rm
d}^{2n+1}(\mu_{\rm d}^{\rm i};\epsilon)$ of the particle population
has escaped downstream and accumulates to form the outgoing particle
population $f_{\rm out}(\epsilon)=\sum_{n=0}^{n=+\infty} f_{\rm
out}^{2n+1}(\epsilon)$. By following each shock crossing, and using
Eqs.~(\ref{eq:mapud_f}),(\ref{eq:mapud_L}),(\ref{eq:mapdu_f}),
(\ref{eq:mapdu_L}) one can follow the evolution of $f_{\rm d}$,
$f_{\rm u}$ and $f_{\rm out}$, starting from a mono-energetic and
isotropic initial injection distribution upstream. The distribution
$f_{\rm out}(\epsilon)$ eventually provides the accelerated particle
population. A similar formal development of the acceleration process
by repeated shock crossings has been proposed independently by Vietri
(2002): the flux of particles crossing the shock in the stationary
regime, noted $J_{\rm in}$ in Vietri (2002) is related to the above as
$J_{\rm in} = C \sum_{n=0}^{n=+\infty} f_{\rm d}^{2n+1}$ with $C$ a
normalization constant.

This method assumes that the angular probability laws do not depend on
rigidity. This is true in the diffusive limit but one
might expect some weak dependence to appear in the relativistic limit
$\Gamma_{\rm s}\gg 1$. Indeed we have found numerically such a weak
dependence of ${\cal P}_{\rm d}$ and ${\cal P}_{\rm u}$ on the
particle rigidity. However it remains weak, and the change in ${\cal
P}$ averages to a few percent when the rigidity changes by two orders
of magnitude. In terms of energy spectral index $s$, this dependence
introduces an error of $\delta s=\pm 0.02$ for $\Gamma_{\rm s}=2$ up to
$\delta s=\pm 0.04$ for $\Gamma_{\rm s}=100$. Therefore in the
following we neglect the dependence on rigidity but keep the above
errors as uncertainties on our results. Note that one can in principle
incorporate this dependence on rigidity in our method at the expense
of having to calculate the probability laws ${\cal P}$ for a wide
range of values of the rigidity.

The present technique has significant advantages when compared to
standard Monte-Carlo techniques which follow the particle trajectories
on both sides of the shock through the whole acceleration
process. Indeed, provided one neglects the dependence on rigidity of
${\cal P}$, one can simulate the trajectories of particles of high
rigidity only (near the end of the resonance range) which are must
faster to integrate than the trajectories of particles of small
rigidity since the ratio of scattering time to Larmor time decreases
with increasing rigidity. The direct Monte-Carlo methods also suffer
from the problem of a small dynamic range of the magnetic fields as
compared to the wide dynamic range of particles momenta. The present
method also offers a significant gain in signal as will be obvious in
the following. Finally our method differs from Gallant et al. (2000)
as they use It\^o differential techniques to simulate scattering
downstream and analytical methods for scattering in a regular magnetic
field upstream assuming $\Gamma_{\rm s}\gg1$. In contrast, the present
simulations can be applied to any shock Lorentz factor and magnetic
field configuration. Furthermore they use Monte-Carlo methods to
simulate the acceleration process after constructing the laws of
return probability while we directly fold over repeatedly the
probability distributions in conjunction with the energy gain formula.

\section{Results}

The downstream return probability to the shock as a function of
ingress pitch angle cosine is shown in Fig.~\ref{fig1} for
$\Gamma_{\rm s}=2,100$. The return probability for $\Gamma_{\rm
s}=100$ is an asymptote which is reached to within a percent as early
as $\Gamma_{\rm s}=5$.  In Fig.~\ref{fig4}, we show the average energy
gain $\langle g\rangle\equiv \langle \epsilon_f\rangle /\langle
\epsilon_i\rangle$ per cycle $u\rightarrow d \rightarrow u$ (diamonds)
and $d\rightarrow u \rightarrow d$ (triangles) for $\Gamma_{\rm
s}=100$. This energy gain is defined as the ratio of the average
energies at the end $\langle\epsilon_f\rangle$ and at the beginning of
the cycle $\langle\epsilon_i\rangle$, with
$\langle\epsilon\rangle\equiv\int{\rm d}\mu{\rm d}\epsilon \,\epsilon
f(\mu,\epsilon)/\int{\rm d}\mu{\rm d}\epsilon\,f(\mu,\epsilon)$.  The
average energy gain in each cycle subsequent to first shock crossing
is $\simeq1.93$ per cycle u--d--u for $\Gamma_{\rm s}=100$, and this
asymptotic value is reached immediately after the first cycle. This is
a rather dramatic confirmation of the analytical expectations of
Gallant \& Achterberg (1999) and Achterberg et al. (2001) which had
argued that only the first cycle should yield a gain
$\approx\Gamma_{\rm s}^2$ since the anisotropy of the distribution
function upstream is so pronounced that the gain in subsequent cycles
is reduced to $\approx 2$.

An example of the spectrum of accelerated particles for $\Gamma_{\rm
s}=100$ after 20 cycles $u\rightarrow d\rightarrow u$ is shown in
Fig.~\ref{fig3}; the thin lines in this figure show the fractions of
particles $f_{\rm out}^{2n+1}$ that escape after $2n+1$ shock
crossings. One clearly sees in this figure the piling up of
populations of particles of ever decreasing size (due to finite escape
probability) and ever increasing energy which gives rise to the
accelerated population $f_{\rm out}=\sum_n f_{\rm out}^{2n+1}$.  The
spectral index of the escaping population for $\Gamma_{\rm s}=100$ is
here $s=2.26\pm0.04$ (incorporating the error due to dependence of
${\cal P}$ on rigidity), in excellent agreement with previous results
by Bednarz \& Ostrowksi (1998), Kirk et al. (2000) and Achterberg et
al. (2001).

Finally, in Fig.~\ref{fig5}, we give the average return probabilities
(open squares), average asymptotic energy gains (diamonds) and
spectral indices (filled circles) for values of $\Gamma_{\rm s}$
comprised between 2 and 100.  The average return probabilities shown
in this figure are the return probabilities shown in Fig.~\ref{fig1}
weighted by the corresponding asymptotic downstream ingress pitch
angle distribution, i.e.  $\langle P_{\rm ret} \rangle =
\lim_{n\rightarrow +\infty} \int {\cal P}_{\rm ret}\left(\mu_{\rm
d}^{\rm i}\right)f_{\rm d}^n(\mu_{\rm d}^{\rm i}){\rm d}\mu_{\rm
d}^{\rm i} / \int f_{\rm d}^n(\mu_{\rm d}^{\rm i}){\rm d}\mu_{\rm
d}^{\rm i} $.  A naive unweighted average of the return probability
shown in Fig.~\ref{fig1} for $\Gamma_{\rm s}=100$ would give 0.33,
whereas the weighted average gives 0.40: the difference is directly
related to the strong anisotropy at shock crossing.

\vspace{0.3cm}
\centerline{{\vbox{\epsfxsize=8.8cm\epsfbox{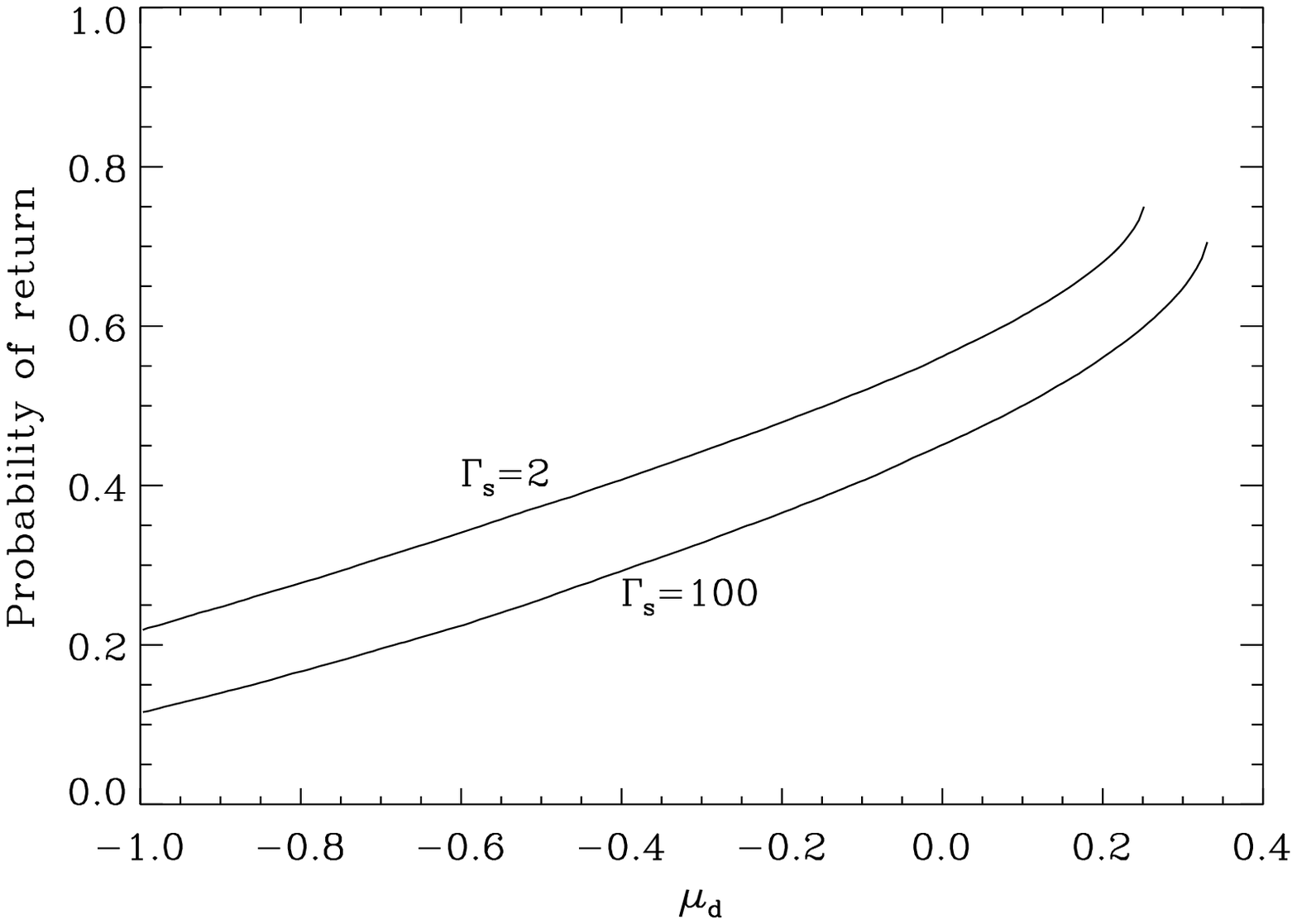}}}}
\figcaption{Probability of return to shock $P_{\rm ret}(\mu_{\rm d})$
downstream as a function of ingress downstream pitch angle cosine
$\mu_{\rm d}$ for $\Gamma_{\rm s}=2,100$. Note that the probability is
defined for $-1\leq \mu_{\rm d}\leq \beta_{\rm s\vert d}$ due to shock
crossing conditions on $\mu_{\rm d}$.\label{fig1}}
\vspace{0.3cm}

\centerline{{\vbox{\epsfxsize=8.8cm\epsfbox{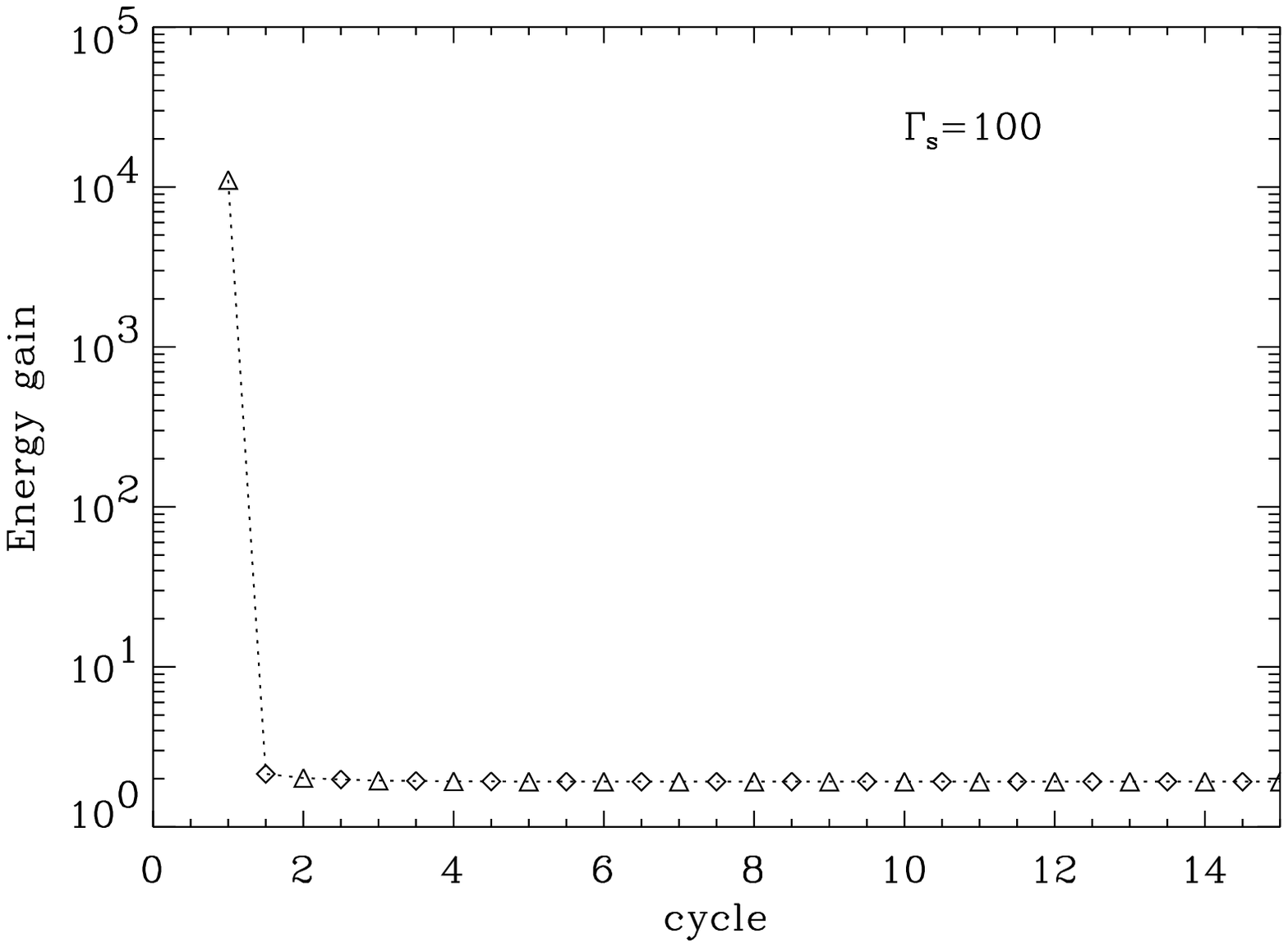}}}}
\figcaption{Averaged energy gain per cycle $u\rightarrow d\rightarrow
u$ (diamonds) and $d\rightarrow u\rightarrow d$ (triangles) for
$\Gamma_{\rm s}=100$. \label{fig4}}
\vspace{0.3cm}

\centerline{{\vbox{\epsfxsize=8.8cm\epsfbox{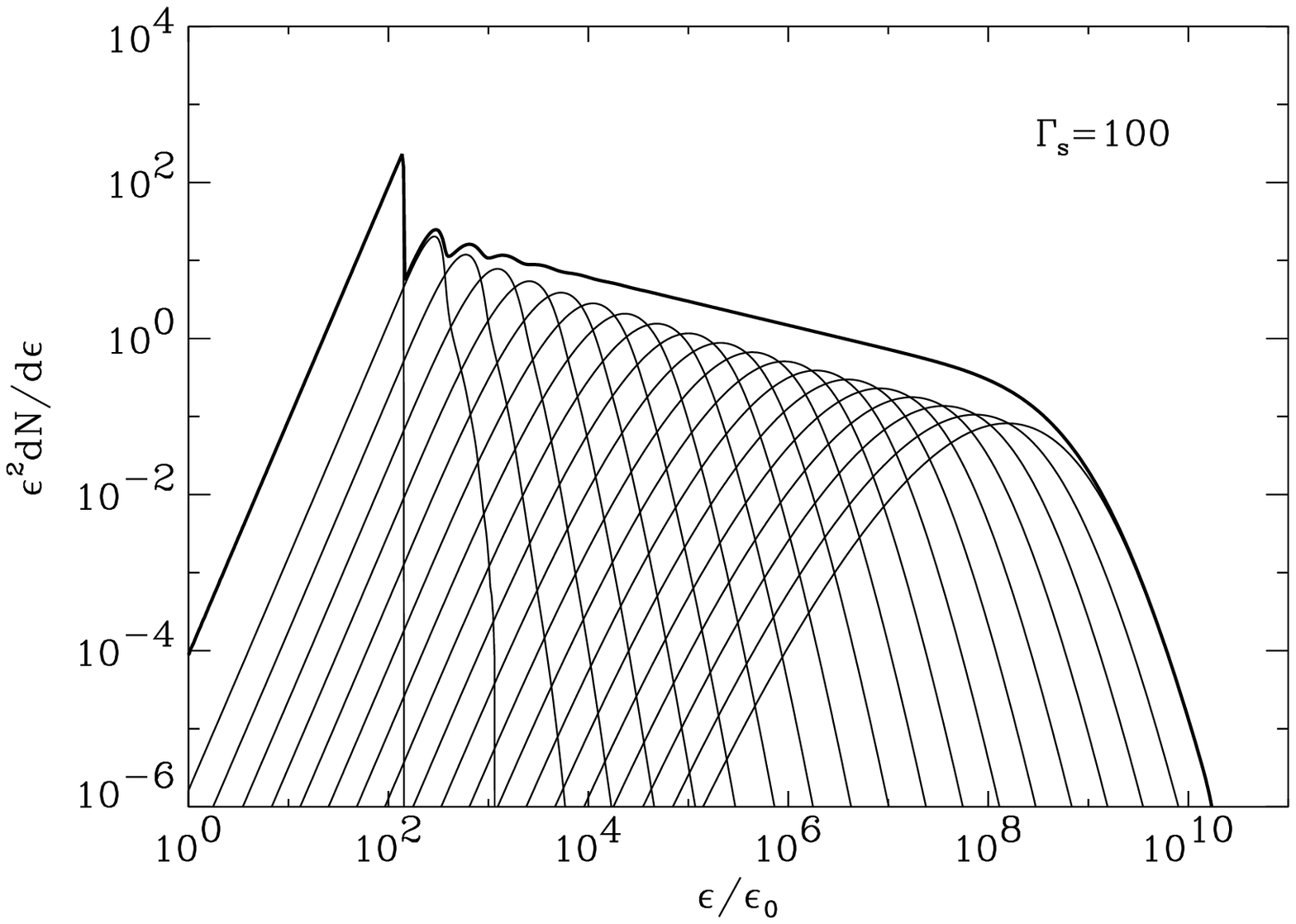}}}}
\figcaption{Spectrum of particles escaping downstream (thick line) as
a function of momentum after 20 cycles for $\Gamma_{\rm s}=100$; in
thin lines, the spectra of particles escaping downstream after each
cycle.  \label{fig3}}
\vspace{0.3cm}

\centerline{{\vbox{\epsfxsize=8.8cm\epsfbox{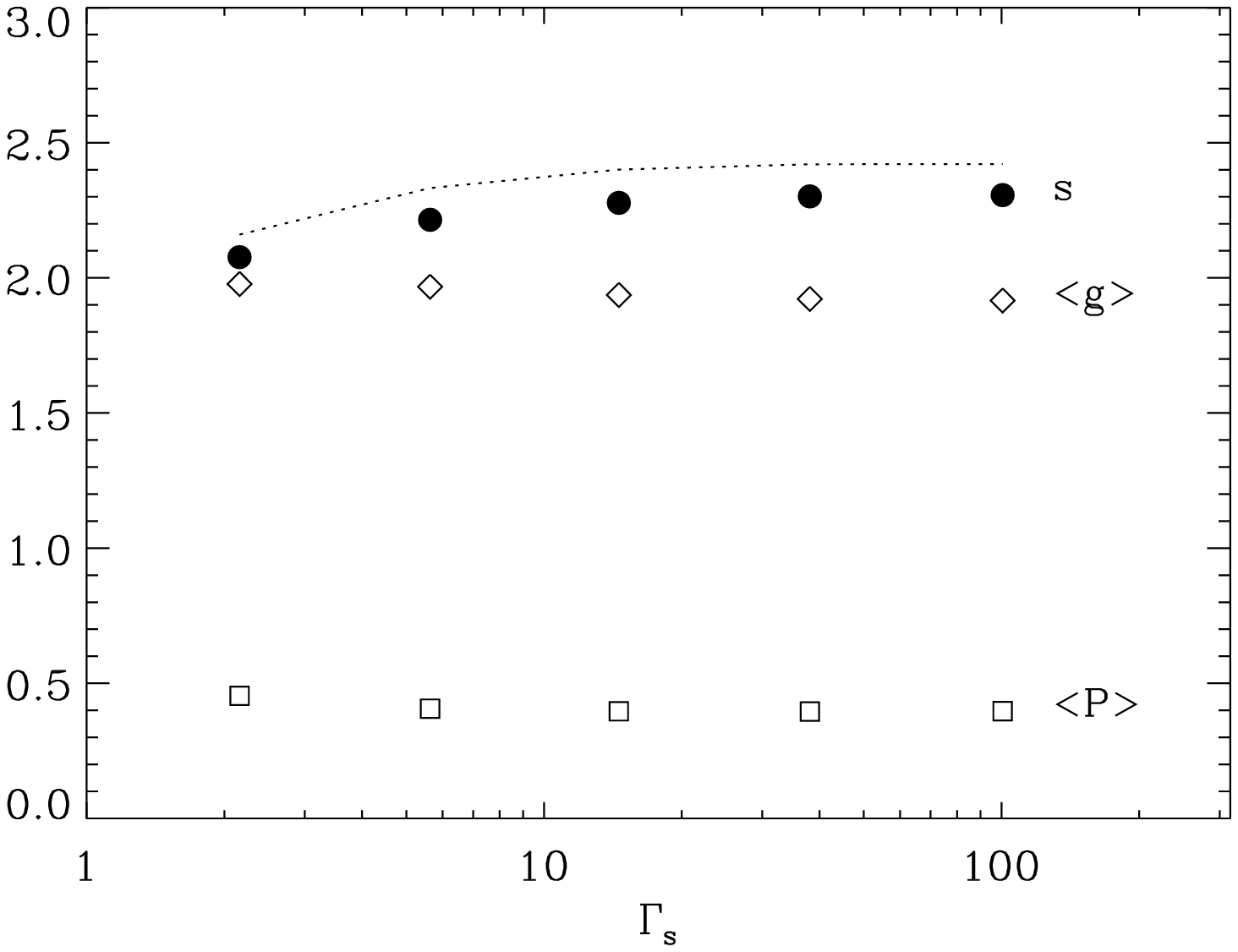}}}}
\figcaption{Average return probabilites (open squares), average energy
gain per cycle (dopen diamonds) and spectral slope (filled circles) as
a function of $\Gamma_{\rm s}$. The dotted line shows the
approximation to the spectral slope given by the Bell formula using
the average return probabilities and energy gains (see
text). \label{fig5}}
\vspace{0.3cm}

The standard non-relativistic formula for the (energy) spectral index
$s$ (Bell 1978), $s=1 - \log\left(\langle P_{\rm ret}\rangle\right)
/\log\left(\langle g\rangle\right) $, with $\langle g\rangle$ the
average energy gain, is in relatively good agreement with the slopes
obtained, provided one uses the weighted average for the return
probability as above, see Fig.~\ref{fig5}. A more general formula
which includes relativistic effects has been proposed by Vietri
(2002): $\langle P_{\rm ret}\rangle\langle g^{s-1}\rangle = 1$. One
can derive this formula and variants of it by using our
Eqs.~(\ref{eq:mapud_f}),(\ref{eq:mapud_L}),(\ref{eq:mapdu_f}),
(\ref{eq:mapdu_L}). For instance, insert Eq.~(\ref{eq:mapdu_f}) into
Eq.~(\ref{eq:mapud_f}) then sum over $n$ (shock crossing number) to go
to the stationary regime and consider an energy range where $\epsilon
\gg \epsilon_0$, $\epsilon_0$ being the maximal injection energy.
There one expects that $\sum_n f_{\rm d}^{2n+1}\propto \epsilon_{\rm
d}^{-s} \phi(\mu_{\rm d})$, i.e. the distribution factorizes out in an
energy power law times a function of pitch angle, and indeed this
property is verified numerically to high accuracy. Then one introduces
the energy gain per cycle $g(\mu_{\rm d}^{\rm e}, \mu_{\rm d}^{\rm i})
= \epsilon_{\rm d} / \tilde \epsilon_{\rm d}$ and integrates over
$\mu_{\rm d}^i$ both sides of the equation. Finally, dividing one side
by the other yields the following relation, which is a variant of the
formula of Vietri (2002):
\begin{equation}
     \int_{-1}^{\beta_{\rm s\vert d}} {\rm d}\mu_{\rm d}^{\rm i}
     \int_{\beta_{\rm s\vert d}}^{1} {\rm d}\mu_{\rm d}^{\rm e}\,
     {\tilde P}_{\rm ret}(\mu_{\rm d}^{\rm e}) \bar {\cal P}_{\rm u}(\mu_{\rm
     d}^{\rm e},\mu_{\rm d}^{\rm i}) g^{s-1}(\mu_{\rm d}^{\rm
     e},\mu_{\rm d}^{\rm i}) = 1 \ .  \label{eq:VIET}
\end{equation}
In this equation, $\bar{\cal P}_{\rm u}(\mu_{\rm d}^{\rm e},\mu_{\rm
d}^{\rm i})$ simply corresponds to ${\cal P}_{\rm u}$ expressed in
terms of downstream pitch angle cosines, and ${\tilde P}_{\rm
ret}(\mu_{\rm d}^{\rm e})\equiv \int {\rm d}\mu_{\rm d}^{\rm i}{\cal
P}_{\rm d}(\mu_{\rm d}^{\rm i};\mu_{\rm d}^{\rm
e})$. Equation~(\ref{eq:VIET}) is indeed verified to within the numerical
noise ($<1$\%).

Our simulations also provide a direct measurement of the acceleration
timescale, which can be defined as the cycle time in the upstream rest
frame when $\Gamma_{\rm s}\gg1$: $t_{\rm acc}(\epsilon)\approx t_{\rm
u\vert u}(\epsilon) + \Gamma_{\rm r} t_{\rm d\vert
d}(\epsilon/\Gamma_{\rm r})$, where $t_{\rm u\vert u}(\epsilon)$ and
$t_{\rm d\vert d}(\epsilon)$ denote the upstream and downstream return
times measured in their respective rest frames for a particle of
energy $\epsilon$. For $t_{\rm d\vert d}$ we find to within the noise
of the simulations: $t_{\rm d\vert d} \simeq 1.5 \beta_{\rm s\vert
d}^{-1} t_{\rm scatt\vert d}$, with $t_{{\rm scatt}\vert d}$ the
scattering time downstream. The scattering time is given as a function
of Larmor time $t_{\rm L}$ in Casse, Lemoine \& Pelletier (2001), and
for pure turbulence one finds: $t_{\rm scatt}/t_{\rm L}\simeq 0.4
\rho^{-2/3}$ for $\rho\lesssim 0.1$ and $t_{\rm scatt}/t_{\rm L}\simeq
2$ for $0.1\lesssim \rho \lesssim 1$. Here $\rho\equiv k_{\rm
min}r_{\rm L}$ denotes the rigidity, with $k_{\rm min}$ the smallest
wavenumber of the magnetic field modes. The above result for $t_{\rm
d\vert d}$ can be understood as follows. In the non-relativistic limit
one derives $t_{\rm d\vert d}\simeq (2/3\beta_{\rm s\vert d}^2)
(1-\langle P_{\rm ret}\rangle)/\langle P_{\rm ret}\rangle$ and we find
$\langle P_{\rm ret}\rangle \sim (1-\beta_{\rm s\vert d})/2$ in the
relativistic limit, which gives $t_{\rm d\vert d}\sim t_{\rm
scatt\vert d}/\beta_{\rm s\vert d}$.

Gallant \& Achterberg (1999) have conjectured $t_{\rm u\vert u}\approx
t_{\rm L\vert u}/\Gamma_{\rm s}$ for $\Gamma_{\rm s}\gg1$,
corresponding to deflection by an angle of order $1/\Gamma_{\rm s}$ in
a regular magnetic field. We confirm this result up to a small
residual dependence on scattering time/rigidity: $t_{\rm u\vert
u}\approx 5 \Gamma_{\rm s}^{-1}\rho_{\rm u}^{-0.15} t_{\rm L\vert u}$
for $\Gamma_{\rm s}\gtrsim 5$. In the mildly relativistic case, for
$\Gamma_{\rm s}=2$, we find $t_{\rm u}\approx t_{\rm scatt\vert u}$,
which indicates that the particles have time to wander before being
caught by the shock in this case.

The final acceleration time depends on how the magnetic field
downstream $B_{\rm d}$ is related to that upstream $B_{\rm u}$
(Gallant \& Achterberg 1999). If one assumes that $B_{\rm d} =
\Gamma_B B_{\rm u}$ with $\Gamma_B\sim \Gamma_{\rm s}$, and the
turbulence is isotropic downstream but the length scale has been
contracted by a factor $\sim \Gamma_B$, i.e. $k_{\rm min\vert d}\sim
\Gamma_B k_{\rm min\vert u}$, one finds, for $\rho_{\rm d\vert
d}\simeq 1$, $t_{\rm acc}\sim (5\Gamma_{\rm s}^{-0.15} + 9)t_{\rm
L\vert u}/\Gamma_{\rm s}$. The acceleration timescale can thus be as
short as a fraction of Larmor time in the ultra-relativistic limit.

This is a most interesting aspect of relativistic Fermi acceleration,
as it implies that the particles can reach the energy confinement
limit $\epsilon_{\rm cl} = Ze\Gamma B r$ when the acceleration is
limited by expansion losses or by the age of the shock wave. Here,
$r={\rm min}(l, t/\kappa c)$, $l$ is the size of the accelerating
region, $t$ the age or losses timescale and $\kappa=t_{\rm acc}/t_{\rm
L} < 1$, all quantities being expressed in the comoving frame. This is
particularly relevant for the generation of ultra-high energy cosmic
rays in relativistic winds such as $\gamma-$ray bursts (Waxman 1995,
Vietri 1995, Gallant \& Achterberg 1999, Gialis \& Pelletier 2003). In
particular our results for $\Gamma_{\rm s}=2$ are of direct relevance
to the acceleration of protons and electrons in internal shocks of
$\gamma-$ray bursts, while the results for $\Gamma_{\rm s}\gg1$ can be
applied directly to the external shock model of $\gamma-$ray bursts.

To sum up, our simulations confirm the results of Bednarz \& Ostrowski
(1998) concerning the spectral index and those of Achterberg et
al. (2001) concerning the spectral index, the acceleration time scale
and energy gains. Very recently, Ellison \& Double (2002) found a
similar index by taking into account the backreaction of cosmic rays
on the shock structure. In the near future, the present method will be
applied to more general magnetic field configurations including
parallel, transverse, subluminal and superluminal shocks.

\acknowledgments We would like to thank Y. Gallant for fruitful
discussions.



%\appendix

%\section{Appendicial material}


%\begin{figure}
%\plottwo{f2a.eps}{f2b.eps}
%\caption{}
%\end{figure}

%% If you are not including electonic art with your submission, you may
%% mark up your captions using the \figcaption command. See the
%% No more than seven \figcaption commands are allowed per page,
%% so if you have more than seven captions, insert a \clearpage

%% Tables should be submitted one per page, so put a \clearpage before

%% Two options are available to the author for producing tables:  the
%% deluxetable environment provided by the AASTeX package or the LaTeX
%% table environment.  Use of deluxetable is preferred.
%%

%% Three table samples follow, two marked up in the deluxetable environment,
%% one marked up as a LaTeX table.

%% In this first example, note that the \tabletypesize{}
%% command has been used to reduce the font size of the table.
%% Note also that the \label command needs to be placed


\begin{thebibliography}{}
\bibitem[]{Aea01} Achterberg, A., Gallant, Y. A., Kirk, J. G., \&
Guthmann, A. W. 2001, \mnras, 328, 393
\bibitem[]{BH92} Ballard, K. R., \& Heavens, A. F. 1992, \mnras, 259,
89
\bibitem[]{B98} Bednarz, J., \& Ostrowski, M.  1998, \prl, 80, 3911.
\bibitem[]{B78} Bell, A. R. 1978, \mnras, 182, 147.
\bibitem[]{B77} Blandford, R., \& McKee, C. 1977, Phys. Fluids, 19,
1130.
\bibitem[]{C01} Casse, F., Lemoine, M., \& Pelletier, G.  2001,
Phys. Rev. D, 65, 023002.
\bibitem[]{E02} Ellison, D., \& Double, G. 2002, Astroparticle
Physics, 18, 213.
\bibitem[]{G99} Gallant, Y. A., \& Achterberg, A.  1999, \mnras, 305,
L6.
\bibitem[]{Gea00} Gallant, Y. A. et al. 2000, in Gamma-Ray
Bursts, eds. R. M. Kippen et al., AIP Conference Series, Vol.526,
p.524 (New-York: American Institute of Physics)
\bibitem[]{G02} Gallant, Y. A. 2002, to appear in Relativistic Flows
in Astrophysics, eds. A. W. Guthmann et al., Lecture Notes in Physics
(Berlin: Springer-Verlag), {\tt arXiv:astro-ph/0201243}
\bibitem[]{GP03} Gialis, D. \& Pelletier, G., Astropart. Phys. in
press, {\tt arXiv:astro-ph/0302231}.
\bibitem[]{KS87} Kirk, J. G., \& Schneider, P. 1987, \apj, 315, 425
\bibitem[]{KD99} Kirk, J., \& Duffy, P. 1999, J. Phys. G, 25, R163.
\bibitem[]{Kea00} Kirk, J., Guthmann, A., Gallant, Y., \& Achterberg, A.
2000, \apj, 542, 235.
\bibitem[]{O93} Ostrowski, M. 1993, \mnras, 264, 248
\bibitem[]{P81} Peacock, J.  1981, \mnras, 196, 135.
\bibitem[]{V95} Vietri, M. 1995, ApJ, 453, 883.
\bibitem[]{V02} Vietri, M.  2002, {\tt arXiv:astro-ph/0212352}.
\bibitem[]{W95} Waxman, E. 1995, Phys. Rev. Lett., 75, 386. 
\end{thebibliography}
\end{document}